\documentclass[conference,letterpaper]{IEEEtran}

\usepackage{cite}
\usepackage[utf8]{inputenc} 
\usepackage[T1]{fontenc}
\usepackage{graphicx}
\usepackage{theorem}
\usepackage[cmex10]{amsmath}
\usepackage{amsfonts}
\ifCLASSOPTIONcompsoc
\usepackage[caption=false,font=normalsize,labelfon
t=sf,textfont=sf]{subfig}
\else
\usepackage[caption=false,font=footnotesize]{subfi
g}
\fi
\usepackage[acronym,shortcuts]{glossaries}
\usepackage{color}
\usepackage{url}
\usepackage{enumerate}
\usepackage{comment}
\usepackage{cleveref}
\usepackage{multirow}
\usepackage{color}

\usepackage{algorithm}
\usepackage[noend]{algpseudocode}
\algdef{SE}[DOWHILE]{Do}{doWhile}{\algorithmicdo}[1]{\algorithmicwhile\ #1}%

\usepackage{color}

\usepackage[acronym,shortcuts]{glossaries}
\newacronym{SDP}{SDP}{Syndrome Decoding Problem}
\newacronym{QC}{QC}{Quasi-Cyclic}
\newacronym{ISD}{ISD}{Information Set Decoding}
\newacronym{DLP}{DLP}{discrete logarithm problem}
\newacronym{IFP}{IFP}{integer factorization problem}
\newacronym{LDGM}{LDGM}{low-density generator matrix}
\newacronym{DOOM}{DOOM}{Decoding One Out of Many}
\newacronym{DFR}{DFR}{decoding failure rate}
\newacronym{QC-LDPC}{QC-LDPC}{quasi-cyclic low-density parity-check}
\newacronym{QC-MDPC}{QC-MDPC}{quasi-cyclic moderate-density parity-check}
\newacronym{QC-LDGM}{QC-LDGM}{quasi-cyclic low-density generator matrix}
\newacronym{LDPC}{LDPC}{low-density parity-check}
\newacronym{LDPCC}{LDPCC}{low-density parity-check convolutional}
\newacronym{NP}{NP}{non-polynomial}
\newacronym{QD}{QD}{quasi-dyadic}
\newacronym{GRS}{GRS}{generalized Reed-Solomon}
\newacronym{AC-LDPC}{AC-LDPC}{array convolutional low-density parity-check}
\newacronym{PDC-LDPC}{PDC-LDPC}{progressive differences convolutional low-density parity-check}
\newacronym{SC-LDPC}{SC-LDPC}{spatially coupled low-density parity-check}
\newacronym{SC-LDPC-CCs}{SC-LDPC-CCs}{spatially coupled low-density parity-check convolutional codes}
\newacronym{SC-LDPC-CC}{SC-LDPC-CC}{spatially coupled low-density parity-check convolutional code}
\newacronym{AWGN}{AWGN}{additive white Gaussian noise}
\newacronym{BF}{BF}{Bit Flipping}
\newacronym{BER}{BER}{bit error rate}
\newacronym{FER}{FER}{codeword error rate}
\newacronym{TUB}{TUB}{truncated union bound}
\newacronym{BPSK}{BPSK}{binary phase shift keying}
\newacronym{SPA-LLR}{SPA-LLR}{sum-product algorithm with log-likelihood ratios}
\newacronym{RTI}{RTI}{regular time-invariant}
\newacronym{RTI-LDPCC}{RTI-LDPCC}{regular time-invariant low-density parity-check convolutional}
\newacronym{CPA}{CPA}{chosen-plaintext attack}
\newacronym{CCA2}{CCA2}{adaptive chosen-ciphertext attack}
\newacronym{SL}{SL}{security level}

\usepackage{algorithm}
\usepackage{algpseudocode}
\algdef{SE}[DOWHILE]{Do}{doWhile}{\algorithmicdo}[1]{\algorithmicwhile\ #1}%

{\theorembodyfont{\rmfamily}
   }
{\theorembodyfont{\rmfamily}
   }
{\theorembodyfont{\rmfamily}
   }
{\theorembodyfont{\rmfamily}
   }
{\theorembodyfont{\rmfamily}
   }
{\theorembodyfont{\rmfamily}
   }
{\theorembodyfont{\rmfamily}
   }

\def\0{\bar{0}}

\makeatletter
\newcommand{\tpmod}[1]{{\@displayfalse\pmod{#1}}}
\makeatother

\definecolor{ps}{rgb}{1,0,0}
\definecolor{cc}{rgb}{0,0,1}

\newcommand{\mb}[1]{#1}

\begin{document}

\title{\mb{Cryptanalysis of a One-Time \\ Code-Based Digital Signature Scheme}} 

\author{\IEEEauthorblockN{Paolo Santini, Marco Baldi, and Franco Chiaraluce}
\IEEEauthorblockA{Dipartimento di Ingegneria dell'Informazione\\
Universit\`a Politecnica delle Marche\\
Ancona, Italy\\
Email: {p.santini@pm.univpm.it}, \{m.baldi, f.chiaraluce\}@univpm.it}}

\maketitle
\begin{abstract}
We consider a one-time digital signature scheme recently proposed by Persichetti and show that a successful key recovery attack can be mounted with limited complexity.
The attack we propose exploits a single signature intercepted by the attacker, and relies on a statistical analysis performed over such a signature, followed by information set decoding.
We assess the attack complexity and show that a full recovery of the secret key can be performed with a work factor that is far below the claimed security level.
The efficiency of the attack is motivated by the sparsity of the signature, which leads to a significant information leakage about the secret key.
\end{abstract}

\section{Introduction}
Code-based cryptosystems, introduced by McEliece in 1978  \cite{McEliece1978}, rely on the hardness of the \ac{SDP}, which has been proven to be NP-complete for general random codes \cite{Berlekamp1978}.
The best \ac{SDP} solvers for general codes, known as \ac{ISD} algorithms, were first introduced by Prange in 1962 \cite{Prange1962} and significantly improved over years (see \cite{Stern1989,Becker2012} and references therein).
However, all current \ac{ISD} algorithms are characterized by an exponential complexity, even when implemented on quantum computers \cite{Bernstein2010}.
Since \ac{SDP} is one of the oldest and most studied hard problems, and no polynomial time solver is currently known, code-based cryptosystems are among the most promising solutions for post-quantum cryptography \cite{NISTreport2016}.

However, designing a secure and efficient  digital signature scheme based on coding theory  is still an open problem.
The main difficulty is represented by the fact that, typically, in these systems the plaintext and ciphertext domains do not coincide.
Therefore, applying decryption on a general string, for example obtained through a hash function, may result in a failure unless special solutions are adopted.
Proposals trying to address this issue have been proven not to be either efficient or secure (or both, in the worst cases).
A clear evidence of the hardness of finding efficient digital signature schemes based on codes is represented by the fact that no proposal of this type is surviving within the National Institute of Standards and Technology (NIST) competition for the standardization of post-quantum primitives \cite{NISTcall2016}.

Historically, the first digital signature scheme based on error correcting codes is the Courtois-Finiasz-Sendrier (CFS) scheme \cite{Courtois2001}, that uses high rate Goppa codes and follows a {\it hash-and-sign} approach.
This scheme is known to be unpractical, since it has some security flaws (high rate Goppa codes can be distinguished from random codes \cite{Faugere2011}) and requires very large public-keys and long signature times. 

In particular, some schemes might suffer from statistical attacks, i.e., procedures that can break the system through the observation of a sufficiently large number of signatures.
In such a case, the attacked systems are reduced to few-signatures schemes or, in the most conservative assumption, to one-time schemes (each key-pair is refreshed after just one signature).
For instance, the BBC$^+$ scheme proposed in \cite{Baldi2013Sig}, which is based on \ac{LDGM} codes, has been cryptanalized in \cite{Phesso2016} with a procedure that allows forging valid signatures after the observation of thousands of signatures, which limits the life of its keypairs \cite{Baldi2018}.
Another recent proposal is Wave \cite{Debris2018}, based on generalized $(U;U+V)$ codes.
A cryptanalysis procedure of Wave based on the statistical analysis of hundreds of signatures has been proposed in \cite{Barreto2018}. 
However, such a procedure has been disproved in \cite{Debris2018Not}, since it is referred to a degraded version of the scheme.


In this paper, we consider a one-time signature scheme that was recently proposed by Persichetti \cite{Persichetti2018}.
Such a scheme is obtained as a modification of Stern's identification protocol \cite{Stern1994}, and relies on \ac{QC} codes, which allow for both compact keys and low computational complexity.
However, as we show afterwards, this scheme suffers from an attack which leads to a full recovery of the secret key and whose complexity is far below the claimed security level.
Our attack is based on a statistical analysis performed on a single signature, combined with an \ac{ISD} algorithm.

Another attack against the same scheme has been independently developed in \cite{Gaborit2018}. 
The attack in \cite{Gaborit2018} is based on \ac{BF} decoding, which has the advantage of being vary fast compared to other decoders.
However, the success probability of \ac{BF} decoding cannot be predicted analytically. Moreover, in case of a decoding failure, it is not possible to perform further randomized attempts of decoding through \ac{BF}.
Differently from \cite{Gaborit2018}, our attack exploits \ac{ISD}, which permits us to obtain a closed-form formula for the average number of iterations and the relevant complexity needed for a successful attack, which depend only on the system parameters. 
So, while the feasibility of the attack in \cite{Gaborit2018} can only be assessed through numerical simulations, we do not rely on simulations: through an theoretical approach, we show that the security of the scheme is reduced to the complexity of an \ac{SDP} instance, which is far below any reasonable security level. In particular, our analysis shows how the security of the system is related to the hardness of solving an \ac{SDP} instance in which the weight of the searched vector is particularly low. Through this approach, we can make general statements about the effectiveness of the attack on modified parameters sets, showing that meaningful security levels cannot be achieved even resorting to extreme choices for the parameters set.


The paper is organized as follows. In Section \ref{sec:notation} we introduce the notation used. In Section \ref{sec:system} we describe the scheme and its design strategy. In Section \ref{sec:attack} we describe our attack procedure and derive estimates of its complexity. Finally, in Section \ref{sec:conclusion} we report some conclusions.

\section{Notation\label{sec:notation}}
We denote as $\mathcal{R}$ the polynomial ring  $\mathbb{F}_2[x]/(x^p+1)$, where $p$ is an integer and $x$ is a symbolic variable.
We use bold letters to denote vectors 
over $\mathcal{R}^2$, in the form $\mathbf{a}(x)=[a_0(x),a_1(x)]$, with $a_i(x)\in\mathcal{R}$.
Each $\mathbf{a}(x)\in\mathcal{R}^2$ can be unambiguously represented as a vector $\mathbf{a}\in\mathbb{F}_2^{2p}$ in the form \[\mathbf{a} = [a_{0,0},a_{0,1},\cdots,a_{0,p-1},a_{1,0},a_{1,1},\cdots,a_{1,p-1}],\]
where $a_{i,j}$ is the $j$-th coefficient of the $i$-th polynomial, $a_i(x)$, in $\mathbf{a}(x)$.
Let $\mathbb{F}_2$ be the binary field. Given a vector $\mathbf{a}$ over $\mathbb{F}_2$, we denote as $\hat{\mathbf{a}}$ the vector obtained by lifting 
its entries over the integer domain $\mathbb{Z}$; the same notation is used for vectors of polynomials.
Operations involving lifted vectors are performed in the integer domain (i.e., $1+1=2$).
Given a polynomial $a(x)$, 
we define its Hamming weight, $\mathrm{wt}\{a(x)\}$, as the number of its non-null coefficients.
For a vector of polynomials $\mathbf{a}(x)$, the Hamming weight corresponds to the sum of the Hamming weights of its elements.
The support of a polynomial $a(x)$, denoted as $\Im\{ a(x) \}$, is the set containing the indexes of the non-null coefficients of $a(x)$.
Clearly, the Hamming weight of a polynomial corresponds to the cardinality of its support.

We denote as $\mathcal{D}_{n,w}$ the uniform distribution of all binary $n$-uples with weight $w$.
Then, the expression $\mathbf{a}\xleftarrow[]{\$}\mathcal{D}_{n,w}$ means that $\mathbf{a}$ is randomly picked among all the elements in $\mathcal{D}_{n,w}$. 
Since the distribution is uniform, each vector of weight $w$ is picked with probability $1/\binom{n}{w}$.
In the following, we consider only the case of $n=2p$.
With some abuse of notation, the expression $\mathbf{a}(x)\xleftarrow[]{\$}\mathcal{D}_{n,w}$ means that $\mathbf{a}(x)$ is randomly picked among all pairs of polynomials having vectors of coefficients in $\mathcal{R}^2$, each with Hamming weight $w$.

\section{System description\label{sec:system}}
The one-time digital signature scheme we are considering is built upon a public polynomial $h(x)$, that is fixed by the protocol. 
We denote as $\texttt{synd}_h:\mathcal{R}^2\rightarrow \mathcal{R}$ the function that takes as input a vector $\mathbf{a}(x)=[a_0(x), a_1(x)]$ and outputs $a_0(x) + a_1(x) h(x)$.
The scheme additionally requires a hash function $\mathcal{H}_{\delta}(b)$ that takes as input $b$ and outputs a weight-$\delta$  polynomial. 
Parameters of the scheme are the integers $p$, $w^{(e)}$, $w^{(y)}$, $w^{(c)}$ (with $w^{(e)},w^{(y)},w^{(c)} \ll p$). 

The key generation is shown in Algorithm \ref{alg:KeyGen}; the signing key (i.e., secret key) is a vector $\mathbf{e}(x) = [e_0(x), e_1(x)]$, such that $\mathrm{wt}\{\mathbf{e}(x)\}=w^{(e)}$.
The verification key (i.e., public key) is
obtained through the application of $\texttt{synd}_h$ on the secret key.
The signature generation and verification are shown, respectively, in Algorithms \ref{alg:SigGen} and \ref{alg:SigVer}.
The signature verification algorithm returns a boolean variable $\bot$ that is false when the signature is valid and true otherwise.

\begin{algorithm}[h]
\caption{Key generation}\label{alg:KeyGen}
\hspace*{\algorithmicindent} \textbf{Input: }integers $p, w^{(e)}, w^{(y)}$\\
\hspace*{\algorithmicindent} \textbf{Output:} signing key $\mathbf{e}(x)$, verification key $s_e(x)$
  \begin{algorithmic}[1]
   \Procedure{}{}
	\State{$\mathbf{e}(x)\xleftarrow[]{\$}\mathcal{D}_{2p,w^{(e)}}$}
	\State{$s_e(x)\gets \texttt{synd}_h\{\mathbf{e}(x)\}$}
	\State{\Return{$\mathbf{e}(x)$, $s_e(x)$}}
\EndProcedure
\end{algorithmic}
\end{algorithm}
\begin{algorithm}[h]
\caption{Signature generation}\label{alg:SigGen}
\hspace*{\algorithmicindent} \textbf{Input: } message $m$, signing key $\mathbf{e}(x)$, integers $w^{(y)},w^{(c)}$\\
\hspace*{\algorithmicindent} \textbf{Output:} signature $\sigma = \{c(x), \mathbf{z}(x)\}$
  \begin{algorithmic}[1]
   \Procedure{}{}
	\State{$\mathbf{y}(x)\xleftarrow[]{\$}\mathcal{D}_{2p,w^{(y)}}$}
    \State{$s_y(x)\gets \mathrm{synd}_h\{\mathbf{y}(x)\}$}
    \State{$c(x)\gets \mathcal{H}_{w^{(c)}}\{[m,s_y(x)]\}$}
    \State{$\mathbf{z}(x)\gets c(x)\mathbf{e}(x)+\mathbf{y}(x)$} 
    \State{$\sigma \gets \{ c(x),\mathbf{z}(x) \}$}
    \State{\Return{$\sigma$}}
\EndProcedure
\end{algorithmic}
\end{algorithm}
\begin{algorithm}[h]
\caption{Signature verification}\label{alg:SigVer}
\hspace*{\algorithmicindent} \textbf{Input: } 
    message $m$, verification key $s_e(x)$, 
    
\hspace*{\algorithmicindent} \textbf{\textcolor{white}{Input: }} signature $\{c(x),\mathbf{z}(x)\}$
    
\hspace*{\algorithmicindent} \textbf{Output:} verification confirmation $\bot$
  \begin{algorithmic}[1]
   \Procedure{}{}
   \State{$\bot\gets 0$}
   \If {$\mathrm{wt}\{\mathbf{z}(x)\}\leq w^{(c)}w^{(e)}+w^{(y)}$}
    \State{$\bot\gets 1$}
    \State{\Return{$\bot$}}\Comment{Signature rejected}
    \EndIf
	\State{$s_z(x)\gets \texttt{synd}_h\{\mathbf{z}(x)\}$}
    \State{$v(x) \gets c(x)s_e(x)+s_z(x)$}
    \State{$c'(x) \gets \mathcal{H}_{\delta}\{[m,v(x)]\}$}
    \If {$c'(x) \neq c(x)$}
    \State{$\bot\gets 1$}
    \State{\Return{$\bot$}}\Comment{Signature rejected}
    \EndIf 
    \State{\Return{$\bot$}}\Comment{Signature accepted}
\EndProcedure
\end{algorithmic}
\end{algorithm}

\subsection{Security analysis\label{sec:security}}
The security of the scheme is based on the hardness of the \ac{SDP} that, in the binary case, is defined as follows.
\vspace{2mm}
\\\textbf{Syndrome Decoding Problem }
\\\textit{Given $\mathbf{H}\in\mathbb{F}_2^{r\times n}$, $\mathbf{s}\in\mathbb{F}_2^{r}$ and $w\in\mathbb{N}$, find $\mathbf{e}\in\mathbb{F}_2^n$ such that $\mathrm{wt}\{\mathbf{e}\}\leq w$ and $\mathbf{H}\mathbf{e}^T = \mathbf{s}$.}\vspace{2mm}\\

The \ac{SDP} is a well-known problem in coding theory, and has been proven to be NP-complete \cite{Berlekamp1978}; in particular, the solution of the \ac{SDP} can be unique only when $w$  does not exceed the Gilbert-Varshamov (GV) distance $d^{(\texttt{GV})}$, that is defined as the greatest integer such that $\sum_{j=0}^{d^{(\texttt{GV})}-2}\binom{n-1}{j} < 2^{n-k}$.
When $w\leq d^{(\texttt{GV})}$, the best solvers for \ac{SDP} are \ac{ISD} algorithms, whose complexity crucially depends on $w$ and on the code rate.

The security of the scheme is based on the fact that the inversion of $\texttt{synd}_h$ requires the solution of an \ac{SDP} instance.
Let $\mathbf{H}$ be the $p\times 2p$ \ac{QC} matrix obtained by concatenating the identity with the circulant matrix having $h(x)$ as first column.
Let $\mathbf{e}$ and $\mathbf{s}_e$ denote, respectively, the vectors associated to the secret and the public key: then, the following relation holds
\begin{equation}
    \mathbf{H}\mathbf{e}^T = \mathbf{s}_e.
\end{equation}
An opponent trying to recover the secret key $\mathbf{e}$ must solve an \ac{SDP} instance; thus, the weight of $\mathbf{e}$ cannot be smaller than some security threshold value.
In the verification procedure, a crucial aspect is represented by the weight of $\mathbf{z}(x)$, which has maximum value equal to $w^{(c)}w^{(e)}+w^{(y)}$.
Indeed, the authenticity of the signature is guaranteed if there is only one vector $\mathbf{z}(x)$ such that
\begin{equation}
   \texttt{synd}_h\{\mathbf{z}(x)\} = c(x)\texttt{synd}_h\{\mathbf{e}(x)\} + \texttt{synd}_h\{\mathbf{y}(x)\},
\end{equation}
since this proves that $\mathbf{z}(x)$ has been computed through the signing key.
Then, a necessary condition for such a vector to be unique is
\begin{equation}
    w^{(c)}w^{(e)}+w^{(y)}\leq d^{(\texttt{GV})}.
\end{equation}
Obviously, the system is fully broken also if the opponent can perform \ac{ISD} on $\texttt{synd}_h\{\mathbf{y}(x)\}=c(x)s_e(x)+s_z(x)$: then, even $w^{(y)}$ cannot be lower than some security threshold value.

Finally, we must take into account that $\mathbf{z}(x)$ is obtained through linear operations involving sparse polynomials, one of them being $c(x)$, which is part of the signature, and is hence public.
In \cite{Persichetti2018}, the possibility of attacks exploiting such facts has been considered; for this reason, the scheme has been proposed only for the one-time signature case.
However, as we show next, the analysis of a single signature, 
combined with an \ac{ISD} algorithm, is enough to recover the secret key. 

\section{An efficient key recovery attack\label{sec:attack}}
We remember that the signature is composed by the pair $\{c(x),\mathbf{z}(x)\}$, with $\mathbf{z}(x) = c(x)\mathbf{e}(x)+\mathbf{y}(x)$.
Let us write
\begin{equation}
    c(x)=\sum_{v\in\Im\{c(x)\}}{x^{v}},
\end{equation}
where $\Im\{c(x)\}$ contains $w^{(c)}\ll p$ distinct integers.

An opponent can compute the polynomials $z_i^{(v)}(x) = x^{-v}z_i(x)$, for $i=0,1$ and for all $v\in\Im\{c(x)\}$; we have
\begin{align}
\label{eq:lifted_z}
    z_i^{(v)}(x) \nonumber & = x^{-v}y_i(x) + x^{-v}c(x)e_i(x) \\
    & = x^{-v}y_i(x) + e_i(x) + \sum_{\begin{smallmatrix}l\in\Im\{c(x)\}\\\nonumber
    l \neq v\end{smallmatrix}}x^{-v+l}e_i(x)\\
    & = y^{(v)}(x) + e_i(x) + \sum_{\begin{smallmatrix}l\in\Im\{c(x)\}\\
    l \neq v\end{smallmatrix}}{e^{(v-l)}_i(x)}.
\end{align}
The opponent can then lift all such polynomials in the integers domain, and compute the sum 
\begin{equation}
    \hat{d}_i(x) = \sum_{j=0}^p{\hat{d}_{i,j}x^j} =  \sum_{v\in\Im\{c(x)\}}{\hat{z}^{(v)}_i(x)},
\end{equation}
for $i = 0,1$.
We expect high coefficients in $\hat{d}_i(x)$ to be associated to ones in $e_i(x)$.
In fact, all polynomials $z_i^{(v)}(x)$ are obtained as the sum of $e_i(x)$ with other sparse polynomials that depend on the shift $x^{-v}$.
Hence, if an entry belongs to the support of a large number of polynomials $z_i^{(v)}(x)$, then it also belongs to the support of $e_i(x)$ with high probability.

The opponent can exploit this fact to estimate the coefficients of $\mathbf{e}(x)$.
In particular, let $\mathbf{e}'(x)=[e_0'(x), e_1'(x)]\in\mathcal{R}^2$ be a vector with coefficients
\begin{equation}
    e'_{i,j}=\begin{cases} 0 &\text{if $\hat{d}_{i,j}<b$,}\\
    1 &\text{if $\hat{d}_{i,j}\geq b$,}
    \end{cases}
\end{equation}
where $b$ is an integer $\leq w^{(c)}$.
The vector $\mathbf{e}'(x)$ represents an estimate of $\mathbf{e}(x)$, whose accuracy depends on the choice of $b$.

The opponent can then compute 
\begin{align}
    s^*(x) \nonumber & = s_e(x)+\texttt{synd}_h\{\mathbf{e}'(x)\}\\\nonumber
    & = \texttt{synd}_h\{\mathbf{e}(x)\} + \texttt{synd}_h\{\mathbf{e}'(x)\}  \\\nonumber
    & = \left[ e_0(x)+e'_0(x)\right] + h(x)\left[ e_1(x) + e_1'(x)\right] \\
    & = \texttt{synd}_h\{\mathbf{e}^*(x)\},
\end{align}
where $\mathbf{e}^*(x) = \mathbf{e}(x) + \mathbf{e}'(x)$.
If $s^*(x)=0$, then $\mathbf{e}'(x) = \mathbf{e}(x)$, otherwise
\ac{ISD} can be used to obtain $\mathbf{e}^*(x)$ from $s^*(x)$, and then the secret key can be recovered as $\mathbf{e}(x) = \mathbf{e}'(x) + \mathbf{e}^*(x)$.

The complexity of the whole attack crucially depends on the weight of $\mathbf{e}^*(x)$, which is related to the accuracy of the estimate $\mathbf{e}'(x)$.
As shown in the next section, for the system we consider it is always possible to choose $b$ such that the weight of $\mathbf{e}^*(x)$ has a high probability of being very small. 

\subsection{Attack complexity\label{sec:complexity}}
Let us denote as $w^{(e)}_i$ and $w^{(y)}_i$ the weights of $e_i(x)$ and $y_i(x)$, respectively.
A specific weights partition is uniquely determined by $w^{(e)}_0$ and $w^{(y)}_0$, as $w_1^{(e)} = w^{(e)}-w^{(e)}_0$ and $w_1^{(y)} = w^{(y)}-w^{(y)}_0$.
The probability to have this partition is
\begin{equation}
    P\left\{w_0^{(e)}, w_0^{(y)} \right\} = \frac{\binom{p}{w^{(e)}_0}\binom{p}{w^{(e)} - w^{(e)}_0}}{\binom{2p}{w^{(e)}}}\frac{\binom{p}{w^{(y)}_0}\binom{p}{w^{(y)} - w^{(y)}_0}}{\binom{2p}{w^{(y)}}}.
\end{equation}
Recall \eqref{eq:lifted_z}, and let us define
\begin{equation}
    \tilde{e}^{(v)}_i(x) = \sum_{\begin{smallmatrix}l\in\Im\{c(x)\}\\l\neq v\end{smallmatrix}}{e_i^{(v-l)}}(x),
\end{equation}
from which $z^{(v)}_i(x) = e_i(x) + y^{(v)}_i(x)+\tilde{e}_i(x)$.
Let $\rho_i^{\texttt{null}}$ be the probability that a particular coefficient in the sum $\tilde{e}_i(x)+y_i^{(v)}(x)$ is null.
We can assume that each $e_i^{(v-l)}(x)$ is a random polynomial with weight $w_i^{(e)}$, and define
\begin{equation}
    \rho_i = \sum_{\begin{smallmatrix}j=0\\j\text{\hspace{1mm} even}\end{smallmatrix}}^{w^{(c)}-1}{\binom{w^{(c)}-1}{j}\left(\frac{w^{(e)}_i}{p}\right)^j\left(1-\frac{w^{(e)}_i}{p}\right)^{w^{(c)}-1-j}},
\end{equation}
such that $\rho_i^{\texttt{null}}$ can be estimated as
\begin{equation}
\rho_i^{\texttt{null}} = \left(1-\rho_i\right) \frac{w^{(y)}_i}{p}+ \rho_i\left(1-\frac{w^{(y)}_i}{p}\right).
\end{equation}
Each null coefficient in $\tilde{e}^{(v)}_i(x)+y^{(v)}_i(x)$ results in a match between $e_i(x)$ and $z^{(v)}_i(x)$; thus, the probability that a set coefficient in $e_i(x)$ is also set in $e'_i(x)$ can be estimated as 
\begin{equation}
\rho^{\texttt{set}}_i = \sum_{j=b}^{w^{(c)}}{\binom{w^{(c)}}{j}\left( \rho_i^{\texttt{null}}\right)^j \left( 1-\rho_i^{\texttt{null}}\right)^{w^{(c)}-j}}.
\end{equation}
Similarly, the probability that a null coefficient in $e_i(x)$ is set in $e'_i(x)$ can be obtained as 
\begin{equation}
\rho^{\neg \texttt{set}}_i = \sum_{j=b}^{w^{(c)}}{\binom{w^{(c)}}{j}\left( 1-\rho^{\texttt{null}}_i\right)^j \left( \rho^{\texttt{null}}_i\right)^{w^{(c)}-j}}.
\end{equation}
Let us denote as $u_i^{\texttt{set}}$ and as $u_i^{\neg \texttt{set}}$ the number of coefficients that are correctly and incorrectly set in $e'_i(x)$; then, we have 
\begin{equation}
    \mathrm{wt}\{e^*_i(x)\} = w^{(e)}_i+u^{\neg\texttt{set}}_i-u^{\texttt{set}}_i.
\end{equation}
Let us define
\begin{align}
    P_i^{\texttt{set}}(u^{\texttt{set}}_i) = & \binom{w^{(e)}_i}{u^{\texttt{set}}_i}\left(\rho^{\texttt{set}}_i\right)^{u^{\texttt{set}}_i}\left(1-\rho^{\texttt{set}}_i\right)^{w^{(e)}_i-u^{\texttt{set}}_i}, \nonumber \\
    P_i^{\neg\texttt{set}}(u^{\neg\texttt{set}}_i) = & \binom{p-w^{(e)}_i}{u^{\neg\texttt{set}}_i}\left(\rho^{\neg\texttt{set}}_i\right)^{u^{\neg\texttt{set}}_i} \cdot \nonumber \\
    & \left(1-\rho^{\neg\texttt{set}}_i\right)^{p-w^{(e)}_i-u^{\neg\texttt{set}}_i}.
\end{align}
The probability that $e^*_i(x)$ has weight $w^{(e^*)}_i$ results in
\begin{equation}
    P_i\left\{w^{(e^*)}_i \right\} = \sum_{u^{\texttt{set}}_i=\max{[0,w^{(e)}_i-w^{(e^*)}_i]}}^{w^{(e)}}{P_i^{\texttt{set}}(u_i^{\texttt{set}})P_i^{\neg\texttt{set}}(u^{\neg\texttt{set}})},
\label{eq:Pi}
\end{equation}
where $u^{\neg\texttt{set}} = w^{(e^*)}_i+u_i^{\texttt{set}}-w^{(e)}_i$.
Let $\delta=\delta_0+\delta_1$, then
\begin{align}
\label{eq:weight_pdf}
    P\left\{\mathrm{wt}(\mathbf{e}^*) = \delta \right\} = \sum_{w^{(e)}_0 = 0}^{w^{(e)}} \nonumber &  \sum_{w^{(y)}_0=0}^{w^{(y)}}P\left\{w_0^{(e)}, w_0^{(y)} \right\}\cdot 
    \\& \cdot \sum_{\delta_0 = 0}^{\delta}{P_0}\{ \delta_0\}P_1\{\delta-\delta_0\}.
\end{align}

Through the probability distribution of $\mathrm{wt}\{ \mathbf{e}^*(x)\}$, we can estimate the effectiveness and the complexity of our cryptanalysis. 
The first part of the attack consists in the computation of $s^*(x)$: since it only involves a limited number of shifts, multiplications and sums, we can neglect the complexity of this step.
If $s^*(x)=0$, then the opponent has already fully recovered the secret key.
In all the other cases, the opponent applies \ac{ISD} on $s^*(x)$, in order to determine the vector $\mathbf{e}^*(x)$, whose weight is unknown and is distributed according to Eq. \eqref{eq:weight_pdf}.
For the sake of simplicity we consider the Lee-Brickell \ac{ISD} algorithm \cite{Lee1988}, which takes as input an integer $j$ and, at each iteration, picks an information set and tests all patterns having a maximum of $j$ ones in the selected positions: an iteration is successful if the selected information set contains a maximum of $j$ errors.
In particular, the complexity of each iteration can be estimated as
\begin{equation}
    C_{\texttt{iter}} = p^3+\sum_{l=0}^j\binom{p}{l}.
\end{equation}
Let $P_{\texttt{iter}}$ denote the probability of success for a single iteration.
Then, we have
\begin{equation}
    P_{\texttt{iter}} = \sum_{\delta=1}^{\bar{w}}{P\{\mathrm{wt}(\mathbf{e}^*) = \delta\} \sum_{l = 0}^{\min\{\delta,j\}}\frac{\binom{\delta}{l}\binom{2p-\delta}{p-l}}{\binom{2p}{p}}},
\end{equation}
where $\bar{w}$ is a sufficiently large integer.
The average complexity of \ac{ISD} can then be estimated as
\begin{equation}
    C_{\texttt{ISD}} = \frac{C_{\texttt{iter}}}{P_{\texttt{iter}}}.
\end{equation} 
As we show next, for all instances proposed in \cite{Persichetti2018} we can determine a value of $b$ for which
$\mathrm{wt}\{\mathbf{e}^*(x)\}=0$ holds with high probability or applying \ac{ISD} on $s^*(x)$ has extremely low complexity.
In particular, these statements are motivated by the fact that, with overwhelming probability, $\mathbf{e}^*(x)$ has an extremely low weight, such that finding it through an \ac{ISD} algorithm requires just a small number of iterations. 

\subsection{Results}
\begin{figure}
    \centering
    \includegraphics[keepaspectratio,width = 9 cm]{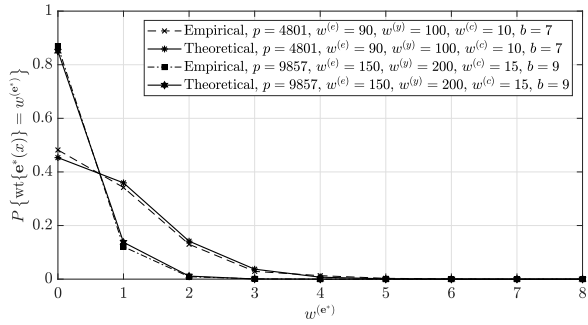}
    \caption{Probability distribution of $\mathrm{wt}\{\mathbf{e}^*(x)\}$. }
    \label{fig:fig_persichetti}
\end{figure}

In Fig. \ref{fig:fig_persichetti} we report the distribution of the weights of $\mathbf{e}^*(x)$ for two instances proposed in \cite{Persichetti2018}.
The empirical distributions have been obtained through numerical simulations on $10,000$ pairs of verification keys and signatures, and have been compared with the theoretical ones expressed by \eqref{eq:weight_pdf}, showing everywhere an excellent agreement.
As we can see, the weight of $\mathbf{e}^*(x)$ assumes very low values with high probability.
This is a clear evidence of the system weakness against the attack. 

In Table \ref{tab:complexity} we have considered the applicability of the attack on the instances proposed in \cite{Persichetti2018}; as we can see, all the instances can be completely broken.
Indeed, $P\{ \mathrm{wt}\{\mathbf{e}^*(x)\}=0\}$ always has high values: thus, with non-negligible probability, the secret key can be fully recovered without invoking \ac{ISD}.
When $\mathrm{wt}\{ \mathbf{e}^*(x)\}>0$, 
it is highly probable that $\mathbf{e}^*(x)$ has an extremely low weight: this results in $P_{\texttt{iter}}$ having very high values, only slightly influenced by the choice of $\bar{w}$ (i.e., choosing $\bar{w}=40$ is already enough to guarantee $P_{\texttt{iter}}\approx 1$). 
This means that the application of \ac{ISD} normally requires a very limited number of operations.
\begin{table}
\caption{Effectiveness of the key recovery attack instances proposed in \cite{Persichetti2018}, for  Lee-Brickell \ac{ISD} with $j=2$.\label{tab:complexity}}
\begin{center}
\begin{tabular}{|cccc|c|c|c|}\hline
$p$ & $w^{(e)}$ & $w^{(y)}$ & $w^{(c)}$ & $b$ & $P\{\mathrm{wt}\{\mathbf{e}^*(x)\}=0\}$ & $C_{\texttt{ISD}}$ \\ \hline\hline
$3072$ & $85$ & $85$ & $7$ & $5$ & $2^{-3.37}$ & $2^{35.10}$\\ \hline
$4801$ & $90$ & $100$ & $10$ & $7$ & $2^{-1.15}$ & $2^{37.58}$\\ \hline
$6272$ & $125$ & $125$ & $10$ & $7$ & $2^{-1.84}$ & $2^{38.37}$\\ \hline
$9857$ & $150$ & $200$ & $15$ & $9$ & $2^{-0.23}$ & $2^{42.54}$\\ \hline
\end{tabular}
\end{center}
\vspace{-0.5em}
\end{table}
\begin{table}
\vspace{-0.5em}
\caption{Effectiveness of the key recovery attack on some system instances with $p=4801$, for  Lee-Brickell \ac{ISD} with $j=2$.\label{tab:extreme_instances}}
\begin{center}
\begin{tabular}{|ccc|c|c|c|}\hline
$w^{(e)}$ & $w^{(y)}$ & $w^{(c)}$ & $b$ & $P\{\mathrm{wt}\{\mathbf{e}^*(x)\}=0\}$ & $C_{\texttt{ISD}}$ \\ \hline\hline
$90$ & $300$ & $8$ & $6$ & $2^{-4.01}$ & $2^{37.05}$\\ \hline
$100$ & $400$ & $6$ & $4$ & $2^{-12.16}$ & $2^{38.95}$\\ \hline
$90$ & $1000$ & $10$ & $7$ & $2^{-13.10}$ & $2^{39.18}$\\ \hline
$90$ & $100$ & $20$ & $12$ & $2^{-0.46}$ & $2^{38.56}$\\ \hline
$180$ & $100$ & $10$ & $7$ & $2^{-16.60}$ & $2^{54.98}$\\ \hline
\end{tabular}
\end{center}
\end{table}

We can also show that changing the system parameters is not enough to significantly raise the security level of the scheme.
In order to give an evidence of this fact, we have considered the case of $p=4801$, for which $d^{(\texttt{GV})}=1058$, and tested different values of $w^{(e)}$, $w^{(y)}$ and $w^{(c)}$.
The results are reported in Table \ref{tab:extreme_instances}.
As we can see, there are no significant changes in the security of the system.
In particular, the last three instances in the table have been designed with a maximum weight of $\mathbf{z}^*(x)$ that is close to $2d^{(\texttt{GV})}$.
This choice is clearly 
extreme since, as explained in Section \ref{sec:system}, this way the uniqueness of the signature is no longer achievable. 
One might think to apply some modifications to the protocol, to  take into account also this possibility in the signature verification algorithm.
However, our results should discourage the attempt.


\section{Conclusion\label{sec:conclusion}}
We have discussed a serious weakness of a recently proposed one-time digital signature scheme.
Our analysis shows that the secret key can be fully recovered with very low complexity, and that changes in the system parameters are not able to restore meaningful security levels.
We point out that, with a few modifications, our attack procedure can be applied to structures different from the \ac{QC} one. This is because it exploits the sparsity of the signature.
As this is an inherent feature of the considered scheme, restoring its security might require deep and structural changes.

\bibliographystyle{IEEEtran}
\bibliography{Archive}

\end{document}